\documentclass[conference]{IEEEtran}
\IEEEoverridecommandlockouts

\usepackage{cite}
\usepackage{threeparttable}
\usepackage{algorithmic}
\usepackage{graphicx} \graphicspath{{images/}}
\usepackage{textcomp}
\usepackage{xcolor}
\usepackage{xspace}
\usepackage{listings}
\usepackage[utf8x]{inputenc} 
\usepackage{caption} 
\usepackage{hyperref}
\hypersetup{
    colorlinks,
    linkcolor={black},
    citecolor={black},
    urlcolor={black}
}
\usepackage{tikz}
\usepackage{balance}

\begin{document}

\newcommand{\todo}[1]{\textbf{\color{red}{#1}}}
\newcommand{\eg}{\textit{e.g.\@}\xspace}
\newcommand{\Eg}{\textit{E.g.\@}\xspace}
\newcommand{\ie}{\textit{i.e.\@}\xspace}


\newcommand{\subsec}[1]{\subsection*{\textbf{#1}}}

\newcommand{\paraheadSpace}{\vspace{0.08in}}

\newcommand{\subsubsec}[1]{\paraheadSpace\subsubsection{#1}}

\def\parahead#1{\paraheadSpace\noindent\textit{#1.}\ }

\newcommand{\nbc}[3]{
	{\colorbox{#3}{\bfseries\sffamily\scriptsize\textcolor{white}{#1}}}
	{\textcolor{#3}{\sf\small \emph{#2}}}
}

\definecolor{skcolor}{RGB}{250,62,159}
\definecolor{bhcolor}{RGB}{62, 229, 250}
\definecolor{slcolor}{RGB}{230,209,57}
\definecolor{kmcolor}{RGB}{165,96,132}
\definecolor{rrcolor}{RGB}{92,119,122}
\newcommand\saketh[1]{\nbc{SRK}{#1}{skcolor}}
\newcommand\brian[1]{\nbc{BH}{#1}{bhcolor}}
\newcommand\sorin[1]{\nbc{SL}{#1}{slcolor}}
\newcommand\kiran[1]{\nbc{KM}{#1}{kmcolor}}
\newcommand\raven[1]{\nbc{RR}{#1}{rrcolor}}

\newcommand{\sysraw}{GUIde}
\newcommand{\longsysraw}{Command Line GUIde}
\newcommand{\sys}{\textsc{\sysraw}\xspace}
\newcommand{\longsys}{\textsc{\longsysraw}\xspace}

\newcommand{\man}{\code{man}\xspace}
\newcommand{\manpage}{\man page\xspace}
\newcommand{\manpages}{\man pages\xspace}

\newcommand{\gl}{\sys-line\xspace}
\newcommand{\gls}{\sys-lines\xspace}

\newcommand{\test}{test case\xspace}
\newcommand{\tests}{test cases\xspace}

\newcommand{\nltobash}{NL2Bash\xspace}

\newcommand{\head}{\code{head}\xspace}
\newcommand{\tail}{\code{tail}\xspace}
\newcommand{\uniq}{\code{uniq}\xspace}
\newcommand{\find}{\code{find}\xspace}
\newcommand{\rsync}{\code{rsync}\xspace}
\newcommand{\split}{\code{split}\xspace}

\newlength{\origcolumnsep}
\newlength{\origintextsep}
\newcommand{\setmarginsforwrapfig}
{
\setlength{\origcolumnsep}{\columnsep}
\setlength{\origintextsep}{\intextsep}
\setlength{\columnsep}{8pt}
\setlength{\intextsep}{5pt}
}
\newcommand{\resetmargins}
{
\setlength{\columnsep}{\origcolumnsep}
\setlength{\intextsep}{\origintextsep}
}

\newcommand{\alignlinetop}[1]{\raisebox{0in}[\dimexpr\height-\intextsep]{#1}}





\definecolor{CodeGrayColor}{RGB}{83, 83, 89}

\newcommand{\codewrap}[1]{\texttt{\textcolor{CodeGrayColor}{#1}}} 
\newcommand{\code}[1]{\mbox{\codewrap{#1}}} 
\newcommand{\mypara}[1]{\paragraph{#1}}
\def\paragraph#1{\noindent\textbf{\textit{#1}.}}

\newcommand{\mytakeaway}[1]{\begin{centering}\begin{tcolorbox}[width=.9\textwidth]#1\end{tcolorbox}\end{centering}}

\definecolor{Pink}{RGB}{208, 2, 208}


 \newcommand*\circled[1]{\tikz[baseline=(char.base)]{
     \node[rounded corners=1.9mm, fill=Pink, inner sep=3pt](char){\textcolor{white}{\textbf{\sffamily{\footnotesize{\textit{#1}}}}}};
 }}



\definecolor{basicColor}{rgb}{0,0,0}
\definecolor{backgroundColor}{rgb}{.97,.97,.97}
\definecolor{identifierColor}{rgb}{0.0,0.0,0.0}
\definecolor{keywordColor}{rgb}{.22,.49,.13}
\definecolor{stringColor}{rgb}{.67,.19,.16}
\definecolor{numberColor}{rgb}{.23,.53,.14}
\definecolor{operatorColor}{rgb}{.61,.18,.96}
\definecolor{commentColor}{rgb}{0.5,0.5,0.5}
\lstdefinestyle{notebookish}{
  frame=single,
  backgroundcolor=\color{backgroundColor}, 
  columns=[l]fullflexible, 
  basicstyle=\linespread{1.0}\footnotesize\ttfamily\color{basicColor}, 
  breaklines=true, 
  keepspaces=true,
  breakatwhitespace=true,
  xleftmargin=1\parindent,
  xrightmargin=1\parindent,
  language=Python,
  showstringspaces=false,
  morekeywords={[1]True,False,None},
  otherkeywords={0,1,2,3,4,5,6,7,8,9,=,+,-,*,/}, 
  morekeywords={[3]0,1,2,3,4,5,6,7,8,9}, 
  morekeywords={[4]=,+,-,*,/}, 
  keywordstyle=\bfseries\color{keywordColor},
  keywordstyle=[2]\color{keywordColor}, 
  keywordstyle=[3]\color{numberColor}, 
  keywordstyle=[4]\bfseries\color{operatorColor},
  commentstyle=\color{commentColor},
  identifierstyle=\color{identifierColor},
  stringstyle=\color{stringColor},
  extendedchars=true,
  literate={∞}{{$\infty$}}1,
  escapeinside={<@}{@>}, 
}
\lstset{style=notebookish}


\renewcommand{\topfraction}{.85}
\renewcommand{\bottomfraction}{.85}
\renewcommand{\textfraction}{.15}
\renewcommand{\floatpagefraction}{.85}
\renewcommand{\dbltopfraction}{.85}
\renewcommand{\dblfloatpagefraction}{.85}


\title{The \longsysraw: \\Graphical Interfaces from Man Pages via AI
\thanks{*Equal contribution.}
}

\author{\IEEEauthorblockN{Saketh Ram Kasibatla\textsuperscript{\raisebox{-1.2pt}{*}},
Kiran Medleri Hiremath\textsuperscript{\raisebox{-1.2pt}{*}},
Raven Rothkopf,
Sorin Lerner,
Haijun Xia,
Brian Hempel}
\{skasibatla,
kmedlerihiremath,
rrothkopf,
lerner,
haijunxia,
bhempel\}@ucsd.edu\\
University of California San Diego,
La Jolla, CA, USA%
}



\IEEEaftertitletext{
\includegraphics[width=\textwidth]{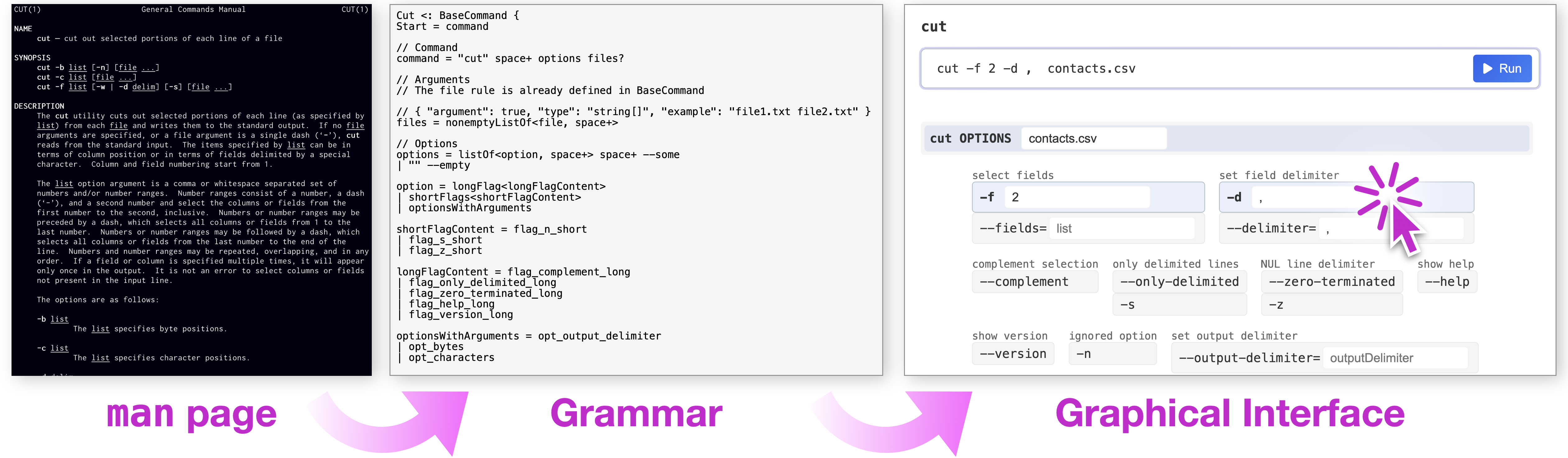}
\label{fig:teaser}
\captionof{figure}{
\sys automatically provides GUI interfaces for command line tools by translating \manpages into specfications.
}
\vspace{1em}
}

\maketitle

\begin{abstract}
Although birthed in the era of teletypes, the command line shell survived the graphical interface revolution of the 1980's and lives on in modern desktop operating systems. The command line provides access to powerful functionality not otherwise exposed on the computer, but requires users to recall textual syntax and carefully scour documentation. In contrast, graphical interfaces let users organically discover and invoke possible actions through widgets and menus. To better expose the power of the command line, we demonstrate a mechanism for automatically creating graphical interfaces for command line tools by translating their documentation (in the form of man pages) into interface specifications via AI. Using these specifications, our user-facing system, called \sys, presents the command options to the user graphically. We evaluate the generated interfaces on a corpus of commands to show to what degree \sys offers thorough graphical interfaces for users' real-world command line tasks.
\end{abstract}


%
%
%
%

\begin{figure*}[t]
	\includegraphics[width=\textwidth]{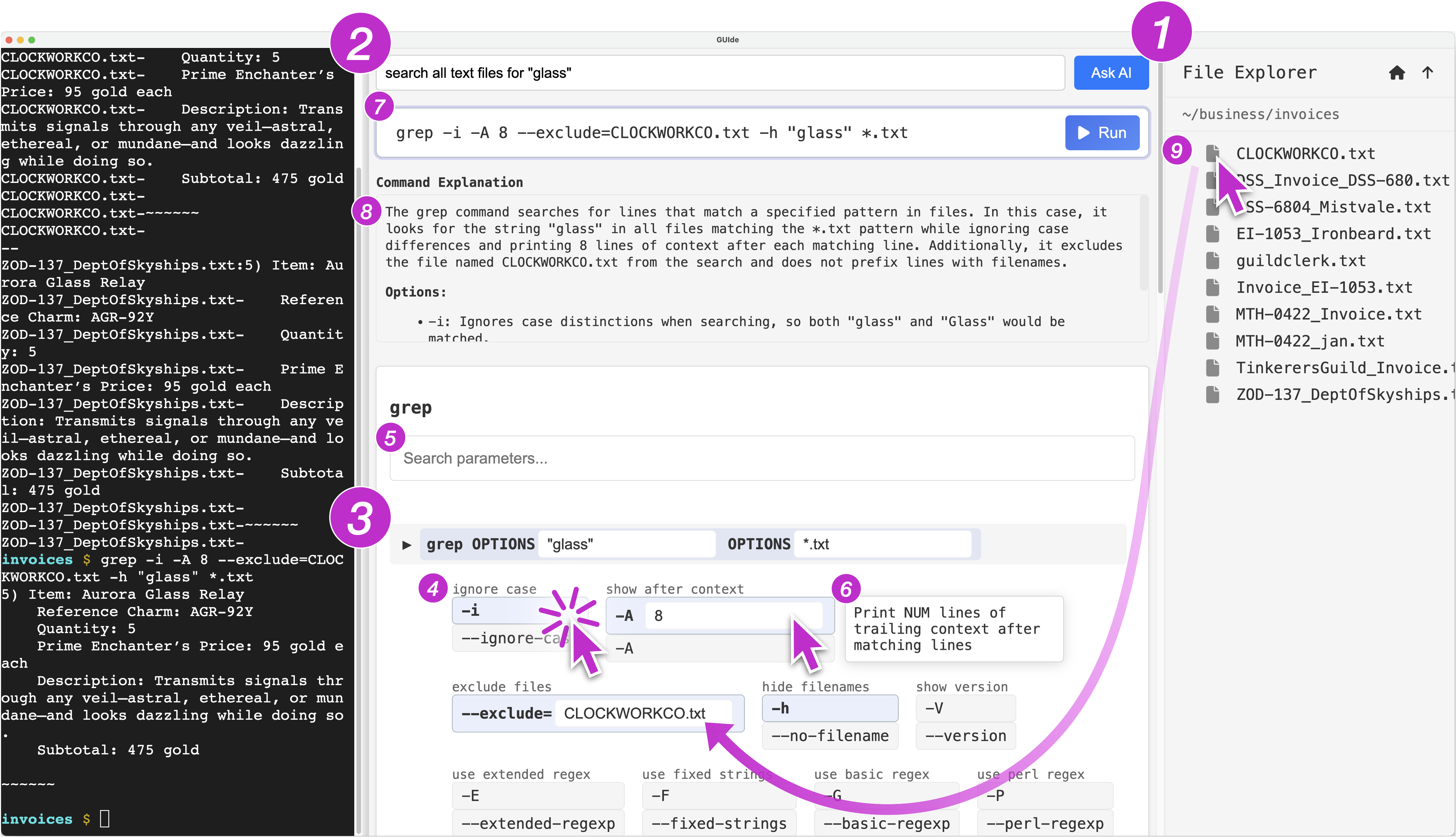}
	\caption{\sys interface, shown for a \code{grep} command.}
	\label{fig:walkthrough}
\end{figure*}

\section{Introduction}

Since at least the dawn of time-sharing mainframes in the early 1960's~\cite{CTSS}, the command line interface (CLI) let users run programs on a computer interactively: type out a command on a keyboard (originally an electric typewriter) and, when completed, see its result (typed back on the paper or shown on an electric display).
Although an interface conceived for teletype machines in an era of glacial computing speeds, the command line interface endured through the graphical personal computer revolution of the 80's and lives on in modern desktop operating systems—Windows, MacOS, and Linux all include a terminal shell—although the command line is no longer considered the normal way to operate a computer.
Nevertheless, programmers still use and write command line tools, in part because building an entire graphical user interface (GUI) for \eg a simple script is a non-trivial undertaking.
But even if writing non-GUI tools is easy for the programmer, using them is hard: one must already know the command name and its flags and argument structure.
Large language model AIs can help users somewhat, allowing them to write their goal in natural language as in recent systems~\cite{Warp},
but AI chat interfaces do not directly help users explore the breadth of all the options of what a command can do. \emph{Discoverability} of options, allowing users to explore, and \emph{interactive tweaking}, allowing users to quickly change properties by clicking, are two key advantages of graphical interfaces not provided by even an AI-augmented command line. Can we ``drag Unix into the 80's''~\cite{Kell2019-ao} by providing graphical interfaces for command line tools, liberating users to quickly explore and modify command options with standard graphical widgets?

\subsec{Related Work}
Several systems have attempted to bridge the command line usability gap, providing graphical front-ends tailored to individual commands. Bespoke~\cite{Bespoke} automatically generates composable GUIs supporting a subset of a command's options by observing user demonstrations. This interaction model lets users flexibly tweak options, but assumes they have the expertise to supply a valid command to begin with, limiting its effectiveness for novice users.
In the early 1990's, Apple's Commando~\cite{Commando} showed graphical dialogs for common Unix commands. These \textit{purpose-built} GUIs streamlined CLI operations, but were not easily adaptable to commands that were unsupported by the system.

Another approach is to provide more \textit{general-purpose} GUIs. PowerShell's Show-Command is one such example~\cite{ShowCommand}. Given a structured specification, Show-Command generates a GUI that constructs and executes any PowerShell command, including user-defined ones. While Show-Command lowers the barrier of entry for novice users, it relies on a structured, machine-readable specification to generate GUIs. Many traditional Unix tools are, however, documented only through natural language \manpages, which would need to be manually translated to a more structured form in order to be used with such a system. 

While each of these systems does provide a GUI to construct commands, they are all \textit{unidirectional}, and only support generating a text command from a GUI. \textit{Bidirectional} interfaces, on the other hand, bridge the gap between graphical and textual programming workflows~\cite{shneiderman1983direct, read1996generating, reiss1984graphical}. These systems let uses modify either the textual or graphical representation of a program, keeping the two in sync~\cite{hempel2019sketch,schreiber2017transmorphic, omar2021filling}.

The advent of large language models (LLMs) also brings great potential for automatic generation of GUIs, especially in light of recent improvements in LLM agents for software engineering~\cite{yang2024swe, xia2024automated, xia2024agentless, liu2024largelanguagemodelbasedagents, Cline}. 
Existing approaches to generating GUIs using LLMs expose options based on user input~\cite{Bespoke,Biscuit,DynaVis, yen2023coladder}. Biscuit~\cite{Biscuit} and DynaVis~\cite{DynaVis} generate GUIs in response to a user prompt, and Bespoke~\cite{Bespoke} does so in response to user demonstrations. Because they are user-driven, they can only display options related to user inputs.Limiting displayed options can reduce cognitive load, presenting users with a clear subset of options, but does so at the expense of discoverability.

\subsec{Our Approach: GUIs from Man Pages via AI}

Instead, we aim for a \emph{maximal} approach, to show an exhaustive list of a command's parameters.
To support both extant and new commands, we aim for \emph{automatic} generation of GUIs from natural language documentation, specifically, from the \manpages (manual pages) usually provided with a command.
We prompt an AI with a \manpage and ask it to output a grammar that describes valid flags and arguments for a command, which is then presented to the user in our graphical application called \textsc{The} \longsys, hereafter \sys.
\sys offers the user a bidirectional interface for authoring a command: the selected options in the GUI updates live as they type a command, and, in the reverse, graphical interaction on the GUI edits the command. This immediate bidirectionally contrasts with \eg PowerShell's Show-Command, which is only a one-shot generator.
We contribute:

\begin{itemize}
  \item GUI inference: generating GUI specifications for command line utilities, given only a \manpage.
  \item \sys: a bidirectional GUI-and-text terminal application for authoring a running commands.
\end{itemize}

\noindent Below, we introduce \sys through an example, detail the GUI inference process, and then critically evaluate the automatically-generated GUIs for 20 common commands.

\section{\longsys Example}

John, a novice command line user, is looking for prior pricing information about an item for his business, but needs to search multiple files at once.\footnote{Mythical invoices for the example are AI-generated (OpenAI \code{gpt-4.1}).}\footnote{Our supplementary materials include an anonymous video of this example.} He knows he can likely do this with command line tools, and he starts up \sys. \sys, shown in  \autoref{fig:walkthrough}, presents a terminal at left, a command editor in the middle, and a file explorer at right.

\parahead{File explorer} The \emph{file explorer} \circled{1} lets John click to navigate to the directory with all his invoices. The terminal runs the needed \code{cd} commands automatically.

\parahead{AI command generation}
\sys provides an AI prompt box \circled{2} for generating and editing commands, similar to recent AI terminal apps~\cite{Warp}. John enters the prompt ``\code{search all text files for "glass"}''.
The AI produces the command \code{grep "glass" *.txt}, but when John runs it there are no results!

\parahead{Flag discoverability and selection} John wonders if \code{grep} is misconfigured. \sys creates a graphical interface \circled{3} for modifying the command without requiring John to look elsewhere for documentation. The interface offers alternative command forms (here hidden behind a disclosure triangle) as well as a comprehensive list of supported flags for \code{grep}. This lets John \emph{discover} relevant flags. Scanning through them,
he notices the \code{-i} flag, labeled ``ignore case'', and wonders if \code{grep} is case-sensitive by default.
John clicks \code{-i} to \emph{toggle} on the \code{-i} flag \circled{4}, adding it to his command. Upon re-running, he now sees the name of the item, ``Aurora Glass Relay'', in two invoices. But, he cannot see the items' prices. \code{grep} only shows the lines that match the search string, but the prices are on nearby lines.

Wondering if there is a way to show surrounding lines, John enters ``line'' in the \emph{parameters search box} \circled{5} and sees an \code{-A} flag labeled ``show after context''. When he hovers his mouse over the flag, a tooltip \circled{6} says ``Print NUM lines of trailing context after matching lines'', which is what he wants. He toggles \code{-A} on, and fills in its input box with 3.

\parahead{Bidirectional editing} After running the command, he sees 3 is not enough lines to show the price. He could change the number in the same input box, but \sys also supports bidirectional editing: the draft command \circled{7} is text-editable. John changes the 3 to an 8 in the full command text, and the GUI below updates to match automatically.
Now when John runs the command, the prices he wants are displayed.



\parahead{Real-time AI Explanation} As John crafts his command, \sys live updates an AI summary \circled{8} of what the command is expected to do. This helps John build confidence that the command will do what he expects, namely, searching for ``glass'', ignoring case, and displaying 8 lines afterward.

\parahead{File drag-and-drop} John realizes his search includes an old invoice that should be excluded. Instead of manually figuring out the exact path to exclude, John toggles on the \code{--exclude} flag and simply drags and drops \circled{9} the unwanted file into the flag's text box, which fills in the text box with the file name. The command is updated with the proper exclusion syntax. John re-runs the command and inspects the output for the latest price he paid for Aurora Glass Relays.



\parahead{Recap} GUIde streamlines the construction of terminal commands with GUI conveniences. A \emph{file explorer} lets users navigate the filesystem and \emph{drag-n-drop files} into command arguments. \sys lists command flags in a graphical interface, facilitating \textit{discoverability}, and simplifying flag selection by offering \emph{search} and quick \emph{click-to-toggle} to add and remove flags. Editing is \emph{bidirectional} when users want to text-edit the full command rather than use the GUI. And \emph{real-time AI explanations} increase user confidence in their command.

\section{Implementation}

\sys's interfaces are generated based on command-specific grammars that describe valid commands. Although developers could write these grammars by hand, we aim to generate GUIs automatically. We prompt an LLM with a command's \manpage to generate the grammars.
Surprisingly, although most commands are simple, we found that na{\"i}vely prompting LLMs produced unusable grammars with many errors. Below we discuss the more involved prompting and repair process we devised to produce usable grammars.

\subsec{Generating \gls from \manpages}
\label{sec:guideline-from-man-page}

The goal is to produce what we call a ``\gl'', a context-free grammar (in Ohm~\cite{Warth2016-pl}) with extra annotations to support GUI rendering. Rules in the grammar can be optionally annotated as representing either a \emph{flag} or an \emph{argument}:

\begin{itemize}
	\item A \emph{flag} is an optional chunk, \eg \code{-a} or \code{--num=10}, that can be toggled on/off in the generated GUI. Whether flags require one, two, or no leading dashes is \emph{not} hard-coded into \sys, it is based on the grammar structure.
	\item An \emph{argument} is a chunk for user input, rendered as an input box in the GUI, \eg the three boxes in \code{cut -d\fbox{,\vphantom{\rule{0pt}{0.6em}}} -f\fbox{2} \fbox{file.csv}} are arguments. Note that \code{-d\fbox{,\vphantom{\rule{0pt}{0.6em}}}} and \code{-f\fbox{2}} are arguments nested \emph{inside} flags.
\end{itemize}


\begin{figure}
    \centering
    \includegraphics[width=0.4\textwidth]{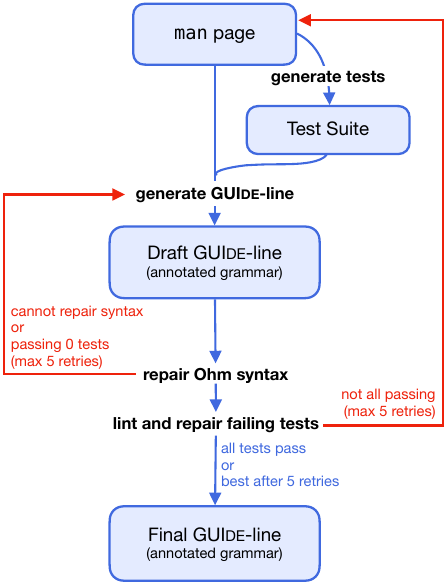}
    \captionof{figure}{
        Automatically creating a \gl from a \texttt{man} page.
    }
    \label{fig:man-page-to-guideline}
\end{figure}


We use a large language model\footnote{\code{claude-3-7-sonnet-20250219} temperature 1 with thinking tokens.} (LLM) to generate \gls using the process outlined in Figure \ref{fig:man-page-to-guideline}.
First, we prompt an LLM to generate a test suite containing valid invocations of a command  based on its \manpage.
We then use the test suite and \manpage to generate a draft \gl.
Finally, we use LLM agents to correct syntactic errors, lint the \gl, and to fix failing test cases. 

\subsec{Test Suite Generation}
\label{sec:generating-test-suite}

To make a test suite, we ask the LLM to generate 10 valid invocations of a command, then to generate a further 10 tests asking it to improve the variety of test cases along several dimensions, including the syntax used to pass arguments, the number of arguments, and use of variables in arguments.
Each of the 20 test cases consists of the text to parse (e.g. ``\code{ls -lah}'') and flags that are expected (e.g. ``\code{-l}'', ``\code{-a}'', and ``\code{-h}''). For a test to pass, the grammar must parse the command successfully \emph{and} the parse tree produced must contain nodes for each of the expected flags. The latter condition causes a test to fail if the \gl contains overly permissive rules that consume more than one flag.

\subsec{Draft \gl Generation}
\label{sec:generate-initial-grammar}

We prompt the LLM to write an annotated Ohm~\cite{Warth2016-pl} grammar, providing (1) the \manpage, (2) the generated test suite, (3) three few-shot \cite{brown2020language} examples detailing ideal output for the \code{ln}, \code{mdfind}, and \code{nl} commands, and (4) several pre-written grammar rules to parse numbers, string literals, embedded commands, flags, etc. This produces an draft annotated grammar (a \gl).

\subsec{Repair with LLM Agents}
\label{sec:editing-with-agents}

\gls from the initial prompt often have errors which make them unusable as-is. We use LLM agents\cite{yang2024swe,Cline,liu2024largelanguagemodelbasedagents,xia2024agentless,xia2024automated} to repair newly written \gls.

An LLM agent is a prompt that is run in a loop that may choose to perform \textit{actions} from a given set. At each step, the LLM is provided the current \gl and error messages, and then performs actions to debug the issue or edit the \gl. The loop terminates after a maximum number of iterations, exiting early if the agent achieves its goal. Each of the repair agents is allowed to perform a subset of the following actions:
\begin{itemize}
    \item \code{replace(diff)} edits the \gl, applying the diff to the first matching search string, as in other software development agents \cite{yang2024swe,Cline}.
    \item \code{read()} returns the current \gl and error message. This helps the agent see the current file state instead of guessing it from the diffs it applied.
    \item \code{parse(example, ruleName)} attempts to parse the string \code{example} under the grammar's \code{ruleName} in the current \gl. This lets an agent debug a problem or test if the problem is fixed.
    \item \code{finish()} marks the task as complete and exits the loop, allowing the agent to decide when it is done.
\end{itemize}

\noindent
We run three agents in series.
First, the \textit{syntax repair agent} fixes syntax errors (\ie invalid Ohm grammars) by using \code{read} and \code{replace}. It is provided with troubleshooting instructions containing suggestions about how to fix the most common syntax errors. The agent may perform up to 10 actions. If the agent cannot produce a valid Ohm grammar, or the grammar it produces passes zero \tests, we regenerate the \gl (maximum of 5 retries).

Next, the \textit{linter agent} repairs \textit{sequencing errors}. These arise because parsing expression grammars~\cite{Ford2004-wx}, like Ohm, parse alternations (\ie \emph{or}-clauses) greedily, always taking the first rule that matches. Thus, longer rules must precede their prefixes in order to be matched (e.g. \code{--print0} must precede \code{--print}). The linter is instructed to look for potential sequencing errors and to fix them, using the \code{parse} action to test rules and \code{replace} to fix them. The linter may perform up to 10 actions, or \code{finish} early.

Finally, the \textit{\test repair agent} fixes failing test cases. As with the linter agent, this agent uses \code{parse} and \code{replace} to debug and fix errors (up to 30 actions). This agent is run once for each failing \test, and is directed to fix the test and to fix instances of the same issue in other parts of the \gl. After each run, the edited \gl is tested against the full test suite, replacing the previous \gl if it has \textit{strictly fewer} failing tests.

If less than 20 of the \tests pass, the \emph{entire} process above starts over with generating new \tests (maximum of 5 retries). Except for unusually complicated commands, no retries are necessary and the above process produces a \gl that can parse all 20 \tests.





\subsec{Bidirectional UI from a \gl}
\label{sec:guideline-to-ui}

When making the \gl, the LLM is instructed to annotate which grammar rules represent flags and arguments. Flag annotations include an identifier (to indicate that \eg \code{-h} and \code{--help} are the same), a short description (displayed in the main UI), and a longer description (displayed in the tooltip). An argument annotation simply marks a rule as an argument, \ie something to be replaced with a text box. The UI displays the grammar rule name as a placeholder in the text box.

To generate the GUI structure, \sys walks the hierarchy of grammar rules and flattens the possible productions into a two level hierarchy: top-level alternatives representing different forms of a command (hidden behind a disclosure triangle in \autoref{fig:walkthrough} \circled{3}), and a set of flags for the alternatives, with equivalent flags grouped together (\eg \code{-V} and \code{--version}). User edits in the GUI serialize the current GUI state to generate the textual command. To provide bidirectional editing, when the user edits the textual command, \sys parses it with the grammar and matches the parse tree nodes with the GUI elements to update the GUI state.

\section{Evaluation}

\begin{table}[t]
  \footnotesize
  \centering
  \begin{minipage}{0.9\textwidth}
    \begin{tabular}{l|r r r r}
      \textbf{Command} & \textbf{\# Recreatable} & \textbf{\# Examples} & \multicolumn{2}{c}{\textbf{Parse Rate}}                        \\ [0.5ex]
      \hline                                                                                                                             \\ [-1.5ex]
      \code{sudo}      & 10                      & 176                  & 100.0\%                                             &                                                 \\
      \code{xargs}     & 9                       & 849                  & 100.0\%                                                                                               \\
      \code{echo}      & 10                      & 344                  & \phantom{0}98.6\%                                                                                     \\
      \code{ssh}       & 10                      & 113                  & \phantom{0}98.2\%                                   &                                                 \\
      \code{mkdir}     & 10                      & 82                   & \phantom{0}97.6\%                                                                                     \\
      \code{cut}       & 10                      & 189                  & \phantom{0}97.4\%                                   &                                                 \\
      \code{tr}        & 10                      & 117                  & \phantom{0}97.4\%                                                                                     \\
      \code{ls}        & 10                      & 107                  & \phantom{0}96.3\%                                                                                     \\
      \code{wc}        & 10                      & 27                   & \phantom{0}96.3\%                                                                                     \\
      \code{grep}      & 8                       & 611                  & \phantom{0}95.7\%                                   &                                                 \\
      \code{dirname}   & 10                      & 64                   & \phantom{0}95.3\%                                                                                     \\
      \code{cat}       & 10                      & 183                  & \phantom{0}95.1\%                                                                                     \\
      \code{tee}       & 10                      & 81                   & \phantom{0}93.8\%                                                                                     \\
      \code{sort}      & 10                      & 188                  & \phantom{0}90.4\%                                                                                     \\
      \code{split}     & 8                       & 78                   & \phantom{0}85.9\%                                   &                                                 \\
      \code{find}      & 0                       & 5162                 & \phantom{0}81.7\%                                   &                                                 \\
      \code{rsync}     & 8                       & 125                  & \phantom{0}80.8\%                                   &          \\ 
      \code{uniq}      & 10                      & 22                   & \phantom{0}77.4\%\\
      \code{tail}      & 7                       & 65                   & \phantom{0}64.6\%                                   & (96.9\%) \\ 
      \code{head}      & 10                      & 70                   & \phantom{0}52.9\%                                   & (97.1\%) \\ [0.5ex] 
      \hline                                                                                                                             \\ [-1.5ex]
      \textit{Mean}    & \textit{9.0}            &                      & \textit{89.8\%} &\\
      \textit{Total}   &                         & \textit{8653}        & &
    \end{tabular}
  \end{minipage}
  \caption{Evaluation metrics for 20 common commands. Parse rates in (parens) are for manually repaired \gls.}
  \label{tab:command-metrics}
\end{table}

To evaluate the automatically-generated \gls and their UIs,
we tested each \gl on two metrics---\textit{parseability}, which tests the grammar itself; and \textit{recreatability}, which tests the whether the generated UI is sufficient to invoke a desired command.
For our corpus, we use the \nltobash\cite{lin2018nl2bash} dataset, which contains bash commands scraped from various online sources.
We generated \gls for the 20 most commonly occurring commands in \nltobash listed in Table \ref{tab:command-metrics} (after splitting pipes into separate commands; we also removed I/O redirects and environment variables). We used \manpages from GNU coreutils for applicable commands, and use \manpages from  Ubuntu 22.04.5 LTS for the remaining commands (\code{grep}, \code{rsync}, \code{ssh}, \code{sudo}, and \code{xargs}).

\parahead{Parseability}
A test command is \textit{parseable} if the \gl parses it successfully. ``Parse Rate'' in Table \ref{tab:command-metrics} is the percentage of parseable invocations. We deduplicate repeated invocations.

Most \gls had a parse rate of over 90\%. However, 6 commands fell below this mark.
\head and \tail used flags not noted in their \manpages. Both support a numerical flag (e.g. \code{-8}) to specify the number of lines to show. As this flag is well-known, it merits being manually added to the \gls. Doing so brings their parse rates above 90\%. 
\find's grammatical structure is highly complex, as it allows for writing nested expressions and boolean logic with its flags.
The LLM struggled with \rsync's exceptionally long \manpage.
%
\split's \gl had mistakes relating to 2 specific flags, accidentally requiring \code{--lines} to take a numerical argument when it could also take a variable, and \code{--filter} mistakenly consumes all flags came after it.
\uniq's 5 failing examples contained uncommon shorthands and argument-passing formats not described in its \manpage.

\parahead{Recreatability}
We consider an invocation \textit{recreatable} if an equivalent command can be reproduced solely by interacting with the graphical interface (\autoref{fig:walkthrough} \circled{3}), without typing in the full command text box (\autoref{fig:walkthrough} \circled{7}), other than typing the command name to pull up the appropriate GUI. For each command, we randomly sampled 10 of the parsable invocations and clicked flags and filled in arguments in the GUI to attempt to re-create an equivalent invocation. ``\# Recreatable'' in Table \ref{tab:command-metrics} shows the number of successes.

Most simple commands are recreatable. Some commands like \code{rsync -rvv} are not supported because the UI only supports using a flag once. The LLM-generated grammar for \code{tail} conflated flags and positional arguments, which was fine for parsing but confused the GUI generator so that user could not supply \emph{both} flags and a file name argument in the same invocation. The worst GUI was for \code{find}: its complicated query syntax was represented by a complicated grammar which the GUI generator interpreted as over 1000 top-level command forms, resulting in an unusable UI.

%
%
%
%

\section{Discussion}

\sys works well for ordinary, simple commands. By using arbitrary grammars, we support, \eg, the user typing \code{ls -lah} and these short flags correctly toggling the \code{-l}, \code{-a}, and \code{-h} flags in the GUI.
Nevertheless, \sys currently fails for commands that cross some threshold of complexity, such as \code{find} which embeds its own query language.

Using arbitrary grammars highlighted to us that the generated GUI is some level of flattening along a continuum. At one extreme, each GUI element would correspond precisely to one grammar rule, supporting the full grammar but requiring the user to navigate a dizzyingly deep nesting of widgets to find the desired options. At the other extreme, the grammar might be fully concretized into all of its (infinitely many!) possible string productions and the user picks from a (infinitely long!) flat list of possible commands.
We chose some midpoint on that continuum and found that, while supporting conventional command structures, our midpoint did not scale to \eg \code{find}.

One strategy for improvement may be to incorporate an additional feedback loop in the automatic generation process: an evaluation of the grammar-generated GUI. Our automatic feedback only looked at \test parse rate, which can work \emph{against} a nice UI: writing a grammar to support odd edge cases and redundant (but allowed) command forms complicates the grammar, adding rules that lead to extraneous GUI elements.

\emph{In conclusion,} we can generate GUIs from documentation (\manpages) with LLMs, using annotated context-free grammars as the link between text and graphical interface, suggesting the possibility of generating GUIs for other ad-hoc structured text, such as configuration languages, industry-specific data formats, and other bespoke formats.

\begin{center}
    —
\end{center}

\noindent
{\footnotesize
See the companion supplement~\cite{supplement} for a video demo, source code, and further details on AI grammar generation and GUI generation from a grammar.
}

\section*{Acknowledgments}

This work was supported by U.S. National Science Foundation Grants No. 2432644 (\emph{Direct Manipulation for Everyday Programming}) and No. 2107397 (\emph{Human-Centric Program Synthesis}).




\balance

\bibliographystyle{IEEEtran}
\bibliography{references}

\end{document}